# *In-situ* dispersion of electrospun nanofibers in PDMS for fabrication of high strength, transparent nanocomposites

Shital Rokade[1], Prasoon Kumar [1, 2*]

[1]Department of Medical Devices, National Institute of Pharmaceutical Education and Research (NIPER)-Ahmedabad, Palaj, Gandhinagar-382355, Gujarat, India.

[2]Department of Biotechnology and Medical Engineering, National Institute of Technology, Rourkela -769008, Odisha India.

*corresponding author: prasoonkumar1985@gmail.com

## Abstract

The polymer nanocomposites find applications in diverse areas ranging from smart materials to bioengineering. They are developed by dispersion of nanomaterials in a bulk phase of a polymeric material. Although several methods facilitate efficient dispersion of nanomaterials in a bulk polymer matrix to create nanocomposites, majority of them follows "heat, beat and treat" processes. These processes are high energy demanding processes. Moreover, the challenge increases when nanomaterials need to be dispersed in a viscous polymeric material. This results in spatial heterogeneity in the dispersion of nanomaterials, eventually leading to compromised mechanical properties of a nanocomposite. Therefore, in the current work, we propose an in-situ, on-step fabrication process of polydimethylsiloxane (PDMS) nanocomposites. Electrospun polyvinyl alcohol (PVA) nanofibers are homogenously dispersed in a PDMS matrix to create a high strength, transparent PDMS nanocomposite. The homogenous dispersion of nanofibers in PDMS matrix is characterised by scanning electron microscopy (SEM), confocal imaging and rheological studies. Further, the prepared PDMS nanocomposite exhibits improved mechanical strength and comparable optical transparency in comparison to native PDMS. Hence, the fabricated PDMS nanocomposites, being resistant to mechanical stress and optically transparent, will find applications as transdermal patches, flexible electronics, microfluidic devices and others.

# 1. Introduction

The polymeric nanocomposites are polyphasic polymeric system wherein one of the phases are nanofillers like nanoparticles, nanofibers, carbon nanotubes and others depending upon the nature of application [1][2]. These polymeric nanocomposites have tremendous advantages over the conventional composites like requirement of minimal quantity of nanofillers, lightweight, enhanced thermal, mechanical and optical properties [3]. Polymeric nanofibers as nanofillers can be fabricated by variety of techniques like self-assembly, freeze-drying synthesis, drawing technique, phase separation, template synthesis, and electrospinning[4]. The electrospinning generates contiguous nanofibers having a high aspect ratio [5]. Such properties of nanofibers can offer better exposed functional groups and higher interfacial surface area for better adhesion interaction with the base polymer matrix during nanocomposites formation [6].

Currently, nanofiber based nanocomposites are fabricated primarily through solution impregnation, phase inversion, solvent casting, and sandwich hot press methods. For instance, Li et al. fabricated starch-based nanocomposites by solvent casting of corn starch solution mixture that was obtained after the dispersion of cellulose nanofiber and montmorillonite through continuous stirring in corn starch solution [7]. In a yet another method, M. Obaid et al. fabricated nanocomposite through a phase inversion method wherein first, the electrospun PVA-Zirconium oxide nanofibers sheet were dispersed in N-methylpyrrolidone/ N, N-dimethylformamide bisolvent system to generate short broken nanofibers and then added to the polysulfone solution for their homogenous distribution before casting to create a nanocomposite [8]. In either of the above methods, the success of homogenous dispersion relied heavily on the chosen solvent systems and degree of stirring/sonication. Further, the dispersion only happens of the short broken nanofibers that may not be sufficient enough for improving the mechanical properties of the developed nanocomposites.

Neppalli et al. proposed sandwich hot press process wherein they sandwiched electrospun nylon-6 nanofibers between the polycaprolactone sheet using a hot press technique to achieve nanocomposite [9]. Although, the lamellar nanocomposites developed had contiguous nanofibers as compared to the broken ones, yet they exhibited anisotropic mechanical properties in the plane perpendicular to the direction of sandwiching. Further, the crucial parameter that determines the success of these sandwich hot press methods is the huge difference in glass transition temperature ($T_g$) and melting temperature ($T_m$) of the base polymer and the nanofibers polymer. The $T_g$ and Tm of the base polymer should be much below Tg and Tm of nanofibers polymers to ensure better penetration of the melt phase of base polymer in the void spaces of the nanofibers sheet [10]. Hence, the above methods are limited to the choice of polymers and also bring anisotropic mechanical properties. P Kumar et al. suggested the use of polydimethyl siloxane (PDMS) (a liquid prepolymer), instead of a melt polymer, to pour over the electrospun nanofibrous membrane to achieve better penetration of PDMS in void spaces of nanofibers and later heat cure to fabricate nanocomposites [11] [12]. Further, Watanabe et al. attempted to directly deposit nanofibers on PDMS solution during electrospining [10]. However, greater depth of dispersion may be a challenge due to increased thickness of high viscous PDMS polymer. Hence, obtaining a thicker block of PDMS having contiguous nanofibers for isotropic mechanical property may be a challenge. Further, nanofibers sheet when sandwiched between the base polymers, the transparency of the developed nanocomposite is compromised as well as its flexibility that plays an important role fabrication of microfluidic devices and flexible electronics [13].

Therefore, in the current manuscript, we have developed a unique setup that enables homogenous dispersion of nanofibers in a viscous PDMS polymer during electrospinning by avoiding strong entanglements among nanofibers. We demonstrate that nanocomposites formed from the above process has high mechanical strength and transparency that is desirable in several biomedical and engineering applications like microfluidic device, implants, optic fibers, skin patches, and drug delivery system.

## 2. Materials and method:

### 2.1 Materials

Polyvinyl alcohol (PVA) (molecular weight =1,60,000) (Himedia Laboratories Pvt. Ltd), Polydimethylsiloxane(PDMS) SYLGARD® 184 (Dow Corning Inc.), Isopropyl alcohol (Fisher Scientific Inc.), Distilled water from Millipore, Elix, USA, Sodium Fluorescein Dye (Sigma Aldrich Pvt Ltd) and Nitrile rubber (non-conductive) procured from Fisher Scientific.

### 2.2.The set-up for in- situ dispersion of nanofibers in PDMS as viscous polymer

12% wt. /vol. of homogenous PVA solution were prepared by continuous addition of PVA to deionised water while being stirred at 200 rpm over a magnetic stirrer (IKA C-MAG HS 7) at 80° C for 2 hours. Simultaneously, PDMS polymer was prepared by thoroughly mixing the pre-polymer (elastomer) with the curing agent at a ratio of 10:1 and then degassed to remove any trapped air bubble to obtain a transparent liquid. Thereafter, the setup as shown in the Figure 1A was designed for in-situ dispersion of PVA nanofibers in PDMS polymer. The prepared PDMS polymer was poured into the glass petri dish (diameter - 150 mm and height - 15mm) and it act as a substrate for the deposition of nanofibers. The corners of the Petri dish were covered with a non-conductive material i,e nitrile rubber. The prepared PVA solution was loaded in a 5 ml syringe having 22 gauge bevel (needle) and electrospun to fabricate nanofibers through an electrospinning set-up (ESPIN-NANO, Pico Chennai, India). The other optimized operational parameters were distance between electrodes - 25cm, voltage - 22.5 KV and flow rate - 1 ml/hr. Finally, the arrangement was made to ensure continuous dripping of PDMS solution from the top (Figure 1A) into the above mentioned petri dish while electrospinnig is going on, leading to simultaneous mixing of nanofiber and PDMS in the collector substrate. The PVA nanofibers dispersed in PDMS polymer was taken and placed in a hot air oven for 12 hour at 80°C. The samples were taken for SEM imaging and solvent etching process. Similarly, the varying amount of PVA nanofibers were dispersed in PDMS using setup (Figure 1A) by regulating the time of nanofiber deposition (10, 15 and 20 min). A part of the samples (before curing of PDMS nanocomposite) was taken for the rheological studies while the other part was cured in a hot air oven for 12 hour at 80°C to obtain solid PDMS nanocomposite for transparency and mechanical testing.

### 2.3. Characterization of nanofibers dispersion in PDMS

The PVA nanofibers were coated with gold at 10 KV for 80 sec using a gold coater and imaged with field emission scanning electron microscope (FE-SEM) (Carl Zeiss, Germany) for morphological characterization at different magnifications. Similarly, PDMS nanocomposite was dipped in liquid nitrogen for 15 min and thereafter mechanically fractured. The fractured surface of the film and the intact surface of PDMS nanocomposite before and after solvent etching was gold coated and imaged using FE-SEM following the above procedure.

### 2.4. Rheological characterization

The rheological characterization was performed using a parallel plate Rheometer (Anton Paar MCR 72). The rheological characterization of nanocomposite was performed before curing the PDMS nanocomposite. The samples were loaded in a 1mm gap between shaft and plate and the temperature maintain at 25°c. The viscosity measurement carried out with plain PDMS and PDMS contain different time interval of nanofibers.

### 2.5. Dye flow experiments

The PDMS nanocomposite samples were immersed in water with frequent sonication for several days for the removal of dispersed PVA nanofibers. Thereafter, nanofiber etched nanocomposite samples were dipped in a dye solution (fluorescein sodium salt dye dissolved in ethanol) for 3 hours ( Figure 2A). Thereafter, the samples were taken out and surface was dried by wiping with tissue papers. Finally, samples were imaged with confocal microscopy (Leica CTR advanced, Germany) at wavelength of 510 -520 nm at different depth.

### 2.6. Mechanical characterization

The universal testing machine (Tinius Olsen-H5KT) having the load cell of 50N was used to evaluate tensile properties of PDMS nanocomposite as well as of the pristine PDMS matrix. The specimen with the dimension of 25mm x 10mm was prepared and placed between the sample holders. Thereafter strain rate of 5mm/min was applied and load and displacement was recorded for the samples (n =3).

### 2.7. Transparency test

The thermally cured nanocomposite film was analysed for its transparency with the help of UV visible spectrophotometry (UV-1800, Shimadzu, Kyoto, Japan). A fresh nanocomposite film was prepared with a dimension of 25mm X 10mm X 1.3mm and placed in an empty cuvette to be considered as a blank control. Thereafter, the above prepared nanocomposite film samples was placed in the cuvette and was scanned in the wavelength range of 400 to 800nm to observe the light transmittance. Furthermore, transparency of nanocomposite were also analysed in a qualitative way. The nanocomposite was placed over the top of a paper printed with NIPER logo and NIPER letters and imaged to decipher the visibility of NIPER logo and NIPER letter.

### 3. Results and discussion

The proposed set-up of nanocomposite fabrication (Figure 1A) enabled an efficient mixing of nanofibers with the dripping PDMS polymer for the proper dispersion of nanofibers. Owing to the simultaneous ejection of nanofibers and PDMS onto a substrate, efficient dispersion have been achieved. The SEM images of PVA nanofibers suggest smooth surface morphology of nanofibers (Figure 1B). The SEM images of PDMS nanocomposites having dispersed nanofibers is shown in the figure 1C. SEM of partially etched PDMS is shown in figure 1D further confirms the presence of nanofibers in PDMS matrix. The dispersed PVA nanofibers, being water soluble, are slowly etched away paving way for the creation of nanochannels and nanoholes in the PDMS (Figure 1D). The nanofibers embossed their impression in PDMS at the time of etching and create nanochannel network. The fabricated nanofibers are contiguous with an average diameter 270 to 360 nm (± 60.49 nm) and there was not any appreciable change in the diameter of nanofiber after dispersion (Figure 1E). Upon placing these etched samples in a fluorescein dye solution, the solution fills the nanochannel network via capillary action (Figure 2A) [11]. The green fluorescence images of these samples shown in the figure 2A (4) suggest that the nanofibers were homogenously dispersed in the sample. We could observe the homogenous dispersion of the nanofibers in a PDMS matrix of 1-2mm of thickness.

The dispersion of nanofillers in PDMS change its inherent viscosity and flow property [14]. The viscosity of pristine PDMS is 3406.40 mPa.s. It was observed that the viscosity of nanofibers dispersed PDMS get increased as compared to the native PDMS. The changes in the viscosity of nanofibers dispersed PDMS depend upon the concentration and the aspect ratio of nanofibers. As the time interval of nanofiber dispersion exceeds from 10 to 20 min in PDMS during the sample preparation, the concentration of nanofibers also increases which ultimately leads to an increase in the viscosity of the sample (Figure 2B and 2C). The increased viscosity was observed due the inability of PDMS to flow freely in a mesh like structure of nanofibers. Further, the physical interaction of PDMS with high surface area of nanofibers further reduced its flow properties. The pristine PDMS, 10, 15, 20 min samples has viscosity such as 3406.4mpa.s, 7503.4mpa.s, 11338mpa.s, 16567mpa.s respectively. Upon increase in the shear rate, the dispersed nanofibers might have experienced breakage due to the shear stress. Perhaps due to this, the viscosity of the sample decreases with the increasing shear rate. However, viscosity of the samples remained higher than the native PDMS sample due to the presence of broken nanofibers. The rheological studies suggested that plain PDMS and the nanofibers dispersed PDMS shows the shear thinning behaviour. However, the 20 min sample exhibits different pattern of viscosity change. This might be probably due to presence of higher density of dispersed, broken nanofiber which alters the flow behaviour of the PDMS-nanofiber mixture. This results in drastic decline in the viscosity of the sample with increasing shear rate. However, moderate dispersion results in flow ability of nanofibers along with PDMS, eventually resulting in a lower rate of decline in the viscous properties. The very high density of nanofibers dispersion in PDMS may have tendency to form aggregates and clumps, result into decreased flow ability of PDMS and shows enhanced viscosity.

The optical transparency being important for several biomedical applications [15]. The optical transmittance of pristine PDMS is about 90%. The dispersion of nanofibers has an effect on the transparency of pristine PDMS. Liao et al. reveal the feature of nanofibers which was observed to be responsible for change in transparency are concentration and the diameter of nanofibers[16]. The 10 and 15 min sample shows the transmittance about 89%, and 79%, respectively while the 20 min sample shows transparency of 75% due to high concentration of nanofibers (Figure 2D). Even after the dispersion of nanofibers in PDMS, the transparencies of the samples are maintained due to uniform dispersion of nanofibers in PDMS. The high concentration of nanofiller in base polymer form agglomerates and which result in high light scattering and ultimately reduces the transparency of nanocomposite [13]. Figure 2C, clearly depict that nanofibers dispersed sample and plain PDMS do not have significant difference in transparency on visual inspection through naked eye. The clear visibility of the NIPER logo and text that was printed on paper kept underneath samples is a testimony to it (Figure 2E).

The nanocomposites have been reported to show significant improvement in their mechanical properties as compared to native polymer (Figure 3). The stress-strain behaviour of different ananocomposites is shown in the figure 3A. The Young's Modulus (YM) of the pristine PDMS was 0.50 MPa whereas 10, 15- and 20-min dispersed nanocomposites show 1.24 MPa, 1.53 MPa and 1.2 MPa, respectively (Figure 3B). In addition, ultimate tensile strength of samples increases with dispersion of nanofibers but there after decreases with increased concentration of dispersed nanofibers. This might be probably due to increase in the stress nucleation site, leading to fracture. Nevertheless, UTM of PDMS dispersed with nanofibers in 15% and 20% by wt. still exhibits better UTM than prestine PDMS (Figure 3C). The mechanical properties of pristine PDMS are increased after dispersion of nanofibers. The factors responsible for enhancing mechanical properties are size, aspect ratio, loading concentration of nanofibers in PDMS. The small diameter of nanofibers having high surface area ultimately increases interfacial adhesion by hydrogen bonding with PDMS polymer. This would lead to transfer of applied load from PDMS to nanofibers thereby increasing the strength of the nanocomposite. The aspect ratio of nanofibers helps in the formation of network inside the PDMS matrix. Transfer of load from PDMS to nanofibres is enhanced by higher aspect ratio of nanofibres in PDMS which paved the improvement in mechanical strength of nanocomposite by performing the fibers bridging process [17]. The mechanical properties of nanocomposite get compromised when incorporated with the nanofillers having low aspect ratio. The low aspect ratio of nanofillers are unable to transfer load from fibers to matrix[17]. The nanofiber loading concentration in a nanocomposite is important in defining the mechanical strength of the nanocomposites. It can be observed from figure 3C that the 20 min nanofibers dispersed sample shows decreased mechanical strength due to high concentration of nanofibers. The higher concentration of nanofibers forms the agglomerates in composite that create the defective site for transfer of load to PDMS and hence, decreases the mechanical strength of nanocomposite. Futher, the presence of nanofibers decreases the % elongation of the PDMS nanocomposites (Figure 4D). This might be due to differential elongation capacity of nanofibers and PDMS matrix polymer.

## 4. Conclusion

A simple set-up has been proposed to facilitate dispersion of PVA nanofibers in PDMS to fabricate PDMS nanocomposite. The process overcomes the traditional heat, beat, treat process to achieve nanocomposite formation. A homogenous dispersion of nanofibers in a viscous PDMS results in nanocomposites with good transparency and high mechanical strength. The process can be extended to any polymer system to generate nanocomposite without worrying about the Tg and Tm of the polymer system. The PDMS nanocomposite obtained can be employed for several engineering applications that were restricted due to poor mechanical properties of the PDMS.

## Acknowledgement

SR and PK would like to acknowledge NIPER-A for proving financial support to carry out this project.

# Figures

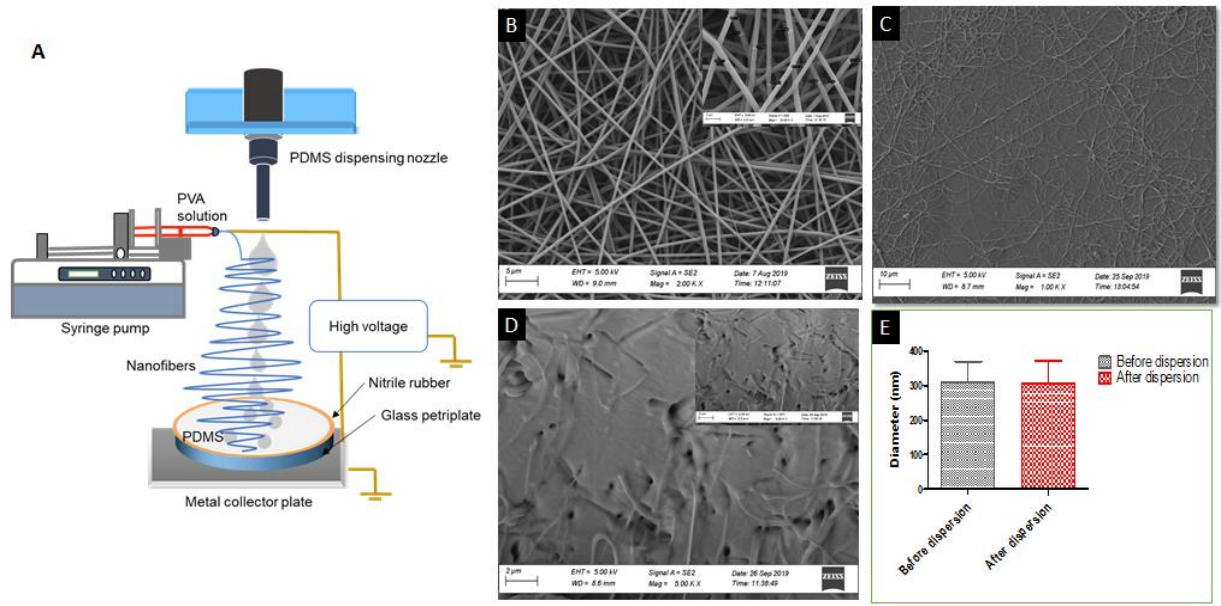

Figure 1 A) Schematic of the experimental set-up for the in-situ fabrication of nanofibers dispersed in PDMS. SEM image of the B) electrospun PVA nanofibers C) electrospun PVA nanofiber dispersed in PDMS D) partially etched PVA nanofibers dispersed in PDMS. E) Graph showing the avg. diameter with standard deviation for PVA fiber before and after dispersion in PDMS

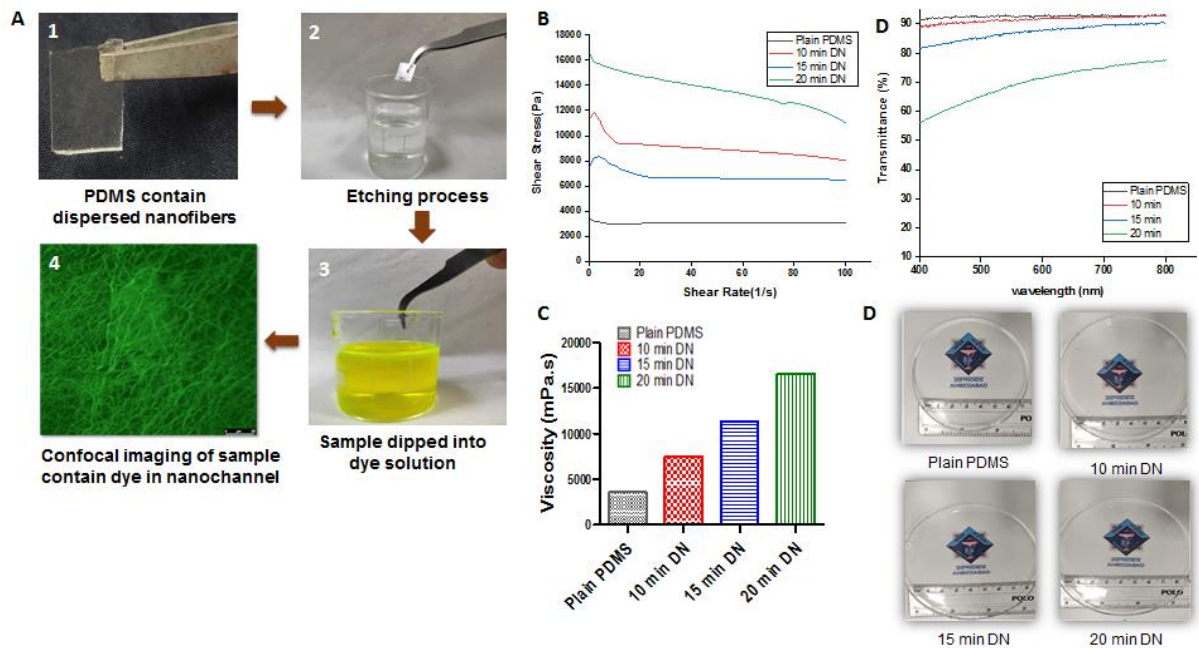

Figure 2 Figure 2 schematic showing the A) process of capillary filling of sodium fluorescein dye in the nanochannels network of PDMS. Graph showing the B) variation of shear stress with shear rate C) viscosity changes D) % transparency in PDMS sample having different density of nanofibers. E) Optical image of NIPER-A logo showing the fabrication of transparent film **(DN- Dispersed Nanofibers)**

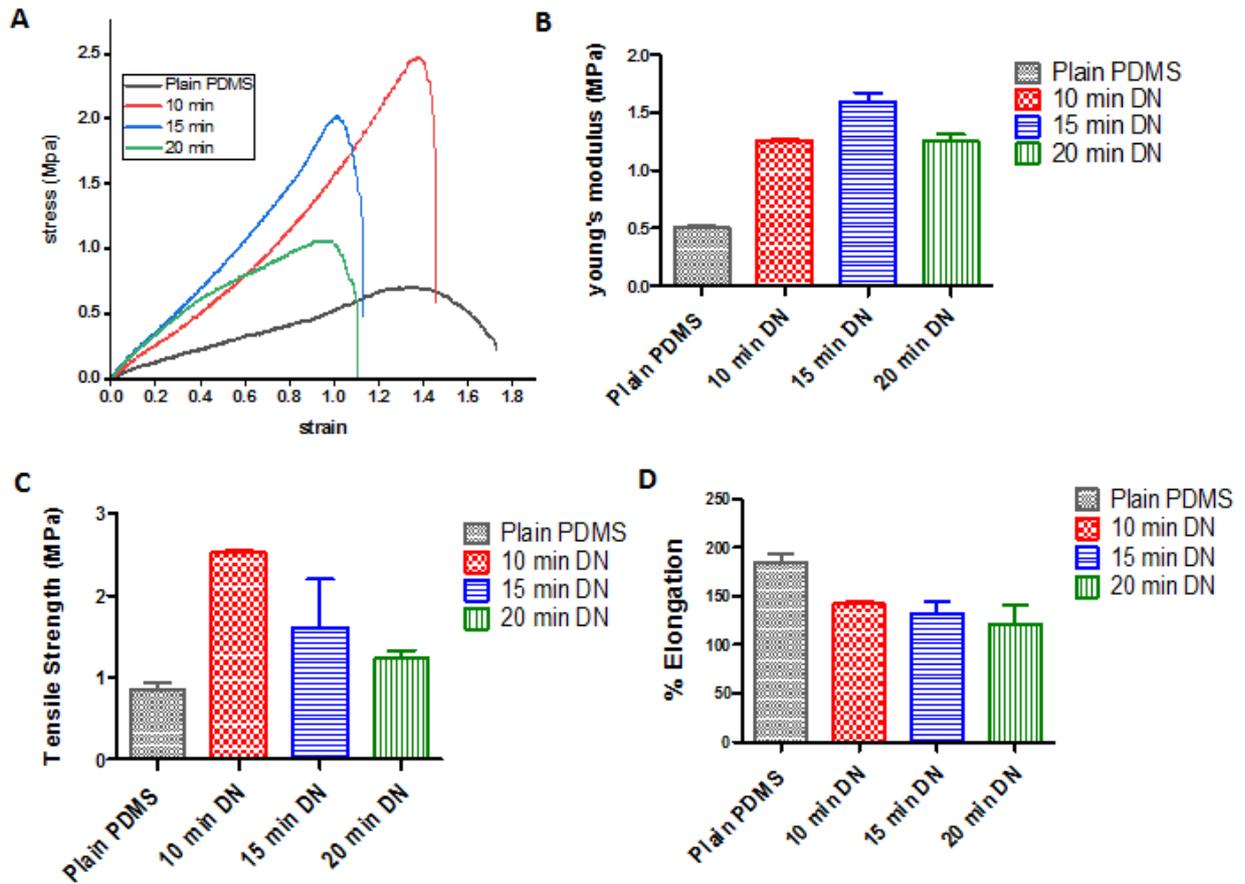

Figure 3 Figure 3 Graph showing the A) Stress Vs. strain B) Young's Modulus C) ultimate tensile strength D) % Elongation in PDMS sample having different density of dispersed nanofibers **(Note: DN- Dispersed Nanofibers)**